\input harvmac
\input epsf
\input amssym
%\draftmode
%
%
\noblackbox
%%%%%%%%%%%%%%%%%%%%%%%%%%%%%%%%%%%%%%%%%%%%%%
%%%%%%%%%%%%%%%%%%%%%%%%%%%
% some stuff needed for figures:
%%%%%%%%%%%%%%%%%%%%%%%%%%%%%%%%%%%%%%%%%%%%%%
%%%%%%%%%%%%%%%%%%%%%%%%%%%
\newcount\figno
\figno=0
\def\fig#1#2#3{
\par\begingroup\parindent=0pt\leftskip=1cm\rightskip=1cm\parindent=0pt
\baselineskip=11pt
\global\advance\figno by 1
\midinsert
\epsfxsize=#3
\centerline{\epsfbox{#2}}
\vskip -21pt
{\bf Fig.\ \the\figno: } #1\par
\endinsert\endgroup\par
}
\def\figlabel#1{\xdef#1{\the\figno}}
\def\encadremath#1{\vbox{\hrule\hbox{\vrule\kern8pt\vbox{\kern8pt
\hbox{$\displaystyle #1$}\kern8pt}
\kern8pt\vrule}\hrule}}
%%%%%%%%%%%%%%%%%%%%%%%%%%%%%%%%%%%%%%%%%%%%%%
%%%%%%%%%%%%%%%%%%%%%%%%%%%
% definitions
%%%%%%%%%%%%%%%%%%%%%%%%%%%%%%%%%%%%%%%%%%%%%%
%%%%%%%%%%%%%%%%%%%%%%%%%%%

\def\ap{\alpha'}

\def\p{\partial}

\def\zb{\overline{z}}

\def\del{\nabla}
\def\eps{\epsilon}

\def\bi{\beta_i}

\def\eps{\epsilon}

\def\bv{\vec{\beta}}
\def\ba{b^{(1)}}

\def\ah{\hat{\alpha}}
\def\ve{\varepsilon}
\def\zb{\overline{z}}

%%%%%%%%%%%%%%%%%%%%%%%%%%%%%%%%%%%%%%%%%%%%%%
%%%%%%%%%%%%%%%%%%%%%%%%%%%
% References

%%%%%%%%%%%%%%%%%%%%%%%%%%%%%%%%%%%%%%%%%%%%%%
%%%%%%%%%%%%%%%%%%%%%%%%%%%

%%%%%%%%%%%%%%%%%%%%%%%%%%%%%%%%%%%%%%%%%%%%%%
%%%%%%%%%%%%%%%%%%%%%%%%%%%
% Title
%%%%%%%%%%%%%%%%%%%%%%%%%%%%%%%%%%%%%%%%%%%%%%
%%%%%%%%%%%%%%%%%%%%%%%%%%%

\Title{\vbox{\baselineskip12pt
%\hbox{hep-th/0508218}
%\hbox{UCLA-05-TEP-XX}
}} {\vbox{\centerline {Nonlinear Magnetohydrodynamics from Gravity  }}} \centerline{James
Hansen\foot{jhansen@physics.ucla.edu} and Per
Kraus\foot{pkraus@ucla.edu}}

\bigskip
\centerline{${}$\it{Department of Physics and Astronomy,
UCLA,}}\centerline{\it{ Los Angeles, CA 90095-1547, USA.}}

\baselineskip15pt

\vskip .3in

\centerline{\bf Abstract}

We apply the recently established connection between nonlinear fluid dynamics and AdS gravity to the
case of the dyonic black brane in AdS$_4$.   This yields the equations of fluid dynamics for a $2+1$ dimensional charged fluid in a background magnetic field.   We construct the gravity solution to second
order in the derivative expansion. From this we find the fluid dynamical stress tensor and  charge current to second and third order in derivatives respectively, along with values for the associated transport coefficients.

%%%
\Date{November, 2008}
%%%%%%%%%%%%%%%%%%%%%%%%%%%%%%%%%%%%%%%%%%%%%%
%%%%%%%%%%%%%%%%%%%%%%%%%%%
% Main text begins here
%%%%%%%%%%%%%%%%%%%%%%%%%%%%%%%%%%%%%%%%%%%%%%
%%%%%%%%%%%%%%%%%%%%%%%%%%%
\baselineskip14pt

%%%%%%%%%%%%%%%%%%%%%%%%%%%%%

%\HansenWU
\lref\Hansen{
  J.~Hansen and P.~Kraus,
  %``Generating charge from diffeomorphisms,''
  JHEP {\bf 0612}, 009 (2006)
  [arXiv:hep-th/0606230].
  %%CITATION = JHEPA,0612,009;%%
}

%\SonSD
\lref\SonSD{
  D.~T.~Son and A.~O.~Starinets,
  %``Minkowski-space correlators in AdS/CFT correspondence: Recipe and
  %applications,''
  JHEP {\bf 0209}, 042 (2002)
  [arXiv:hep-th/0205051].
  %%CITATION = JHEPA,0209,042;%%
}

%\SonVK
\lref\SonVK{
  D.~T.~Son and A.~O.~Starinets,
  ``Viscosity, Black Holes, and Quantum Field Theory,''
  Ann.\ Rev.\ Nucl.\ Part.\ Sci.\  {\bf 57}, 95 (2007)
  [arXiv:0704.0240 [hep-th]].
  %%CITATION = ARNUA,57,95;%%
}

%\BirminghamPJ
\lref\CFTQNN{
  D.~Birmingham, I.~Sachs and S.~N.~Solodukhin,
  ``Conformal field theory interpretation of black hole quasi-normal %%@
modes,''
  Phys.\ Rev.\ Lett.\  {\bf 88}, 151301 (2002)
  [arXiv:hep-th/0112055].
  %%CITATION = PRLTA,88,151301;%%
}

%\cite{Herzog:2002pc}
\lref\SK{
  C.~P.~Herzog and D.~T.~Son,
  ``Schwinger-Keldysh propagators from AdS/CFT correspondence,''
  JHEP {\bf 0303}, 046 (2003)
  [arXiv:hep-th/0212072].
  %%CITATION = JHEPA,0303,046;%%
}

%\HaackXX
\lref\HaackXX{
  M.~Haack and A.~Yarom,
  ``Universality of second order transport coefficients from the gauge-string
  duality,''
  arXiv:0811.1794 [hep-th].
  %%CITATION = ARXIV:0811.1794;%%
}

%\BhattacharyyaJC
\lref\NLFDFG{
  S.~Bhattacharyya, V.~E.~Hubeny, S.~Minwalla and M.~Rangamani,
  ``Nonlinear Fluid Dynamics from Gravity,''
  JHEP {\bf 0802}, 045 (2008)
  [arXiv:0712.2456 [hep-th]].
  %%CITATION = JHEPA,0802,045;%%
}

%\ChamblinTK
\lref\CJ{
  A.~Chamblin, R.~Emparan, C.~V.~Johnson and R.~C.~Myers,
  ``Charged AdS black holes and catastrophic holography,''
  Phys.\ Rev.\  D {\bf 60}, 064018 (1999)
  [arXiv:hep-th/9902170].
  %%CITATION = PHRVA,D60,064018;%%
}

%\KovtunWP
\lref\SH{
  P.~Kovtun, D.~T.~Son and A.~O.~Starinets,
  ``Holography and hydrodynamics: Diffusion on stretched horizons,''
  JHEP {\bf 0310}, 064 (2003)
  [arXiv:hep-th/0309213].
  %%CITATION = JHEPA,0310,064;%%
}

%\KovtunKX
\lref\KovtunKX{
  P.~Kovtun and A.~Ritz,
  ``Universal conductivity and central charges,''
  arXiv:0806.0110 [hep-th].
  %%CITATION = ARXIV:0806.0110;%%
}

%\VanRaamsdonkFP
\lref\VanRaamsdonkFP{
  M.~Van Raamsdonk,
  ``Black Hole Dynamics From Atmospheric Science,''
  JHEP {\bf 0805}, 106 (2008)
  [arXiv:0802.3224 [hep-th]].
  %%CITATION = JHEPA,0805,106;%%
}

%\BhattacharyyaXC
\lref\BhattacharyyaXC{
  S.~Bhattacharyya {\it et al.},
  ``Local Fluid Dynamical Entropy from Gravity,''
  JHEP {\bf 0806}, 055 (2008)
  [arXiv:0803.2526 [hep-th]].
  %%CITATION = JHEPA,0806,055;%%
}

%\BhattacharyyaJI
\lref\BhattacharyyaJI{
  S.~Bhattacharyya, R.~Loganayagam, S.~Minwalla, S.~Nampuri, S.~P.~Trivedi and S.~R.~Wadia,
  ``Forced Fluid Dynamics from Gravity,''
  arXiv:0806.0006 [hep-th].
  %%CITATION = ARXIV:0806.0006;%%
}

%\BhattacharyyaMZ
\lref\BhattacharyyaMZ{
  S.~Bhattacharyya, R.~Loganayagam, I.~Mandal, S.~Minwalla and A.~Sharma,
  ``Conformal Nonlinear Fluid Dynamics from Gravity in Arbitrary Dimensions,''
  arXiv:0809.4272 [hep-th].
  %%CITATION = ARXIV:0809.4272;%%
}

%\BanerjeeTH
\lref\BanerjeeTH{
  N.~Banerjee, J.~Bhattacharya, S.~Bhattacharyya, S.~Dutta, R.~Loganayagam and P.~Surowka,
  ``Hydrodynamics from charged black branes,''
  arXiv:0809.2596 [hep-th].
  %%CITATION = ARXIV:0809.2596;%%
}

%\HaackCP
\lref\HaackCP{
  M.~Haack and A.~Yarom,
  ``Nonlinear viscous hydrodynamics in various dimensions using AdS/CFT,''
  JHEP {\bf 0810}, 063 (2008)
  [arXiv:0806.4602 [hep-th]].
  %%CITATION = JHEPA,0810,063;%%
}

%\ErdmengerRM
\lref\ErdmengerRM{
  J.~Erdmenger, M.~Haack, M.~Kaminski and A.~Yarom,
  ``Fluid dynamics of R-charged black holes,''
  arXiv:0809.2488 [hep-th].
  %%CITATION = ARXIV:0809.2488;%%
}

%\HartnollIP
\lref\HartnollIP{
  S.~A.~Hartnoll and C.~P.~Herzog,
  ``Ohm's Law at strong coupling: S duality and the cyclotron resonance,''
  Phys.\ Rev.\  D {\bf 76}, 106012 (2007)
  [arXiv:0706.3228 [hep-th]].
  %%CITATION = PHRVA,D76,106012;%%
}

%\HartnollIH
\lref\HartnollIH{
  S.~A.~Hartnoll, P.~K.~Kovtun, M.~Muller and S.~Sachdev,
  ``Theory of the Nernst effect near quantum phase transitions in condensed
  matter, and in dyonic black holes,''
  Phys.\ Rev.\  B {\bf 76}, 144502 (2007)
  [arXiv:0706.3215 [cond-mat.str-el]].
  %%CITATION = PHRVA,B76,144502;%%
}

%\HartnollAI
\lref\HartnollAI{
  S.~A.~Hartnoll and P.~Kovtun,
  ``Hall conductivity from dyonic black holes,''
  Phys.\ Rev.\  D {\bf 76}, 066001 (2007)
  [arXiv:0704.1160 [hep-th]].
  %%CITATION = PHRVA,D76,066001;%%
}

%\HerzogIJ
\lref\HerzogIJ{
  C.~P.~Herzog, P.~Kovtun, S.~Sachdev and D.~T.~Son,
  ``Quantum critical transport, duality, and M-theory,''
  Phys.\ Rev.\  D {\bf 75}, 085020 (2007)
  [arXiv:hep-th/0701036].
  %%CITATION = PHRVA,D75,085020;%%
}

%\BalasubramanianRE
\lref\BalasubramanianRE{
  V.~Balasubramanian and P.~Kraus,
  ``A stress tensor for anti-de Sitter gravity,''
  Commun.\ Math.\ Phys.\  {\bf 208}, 413 (1999)
  [arXiv:hep-th/9902121].
  %%CITATION = CMPHA,208,413;%%
}

%\HenningsonGX
\lref\HenningsonGX{
  M.~Henningson and K.~Skenderis,
  ``The holographic Weyl anomaly,''
  JHEP {\bf 9807}, 023 (1998)
  [arXiv:hep-th/9806087].
  %%CITATION = JHEPA,9807,023;%%
}

%\BuchbinderDC
\lref\BuchbinderDC{
  E.~I.~Buchbinder, A.~Buchel and S.~E.~Vazquez,
  ``Sound Waves in (2+1) Dimensional Holographic Magnetic Fluids,''
  arXiv:0810.4094 [hep-th].
  %%CITATION = ARXIV:0810.4094;%%
}

\newsec{Introduction}

Work over the past year has established an elegant connection between long wavelength disturbances
of asymptotically AdS black brane solutions  and relativistic fluid dynamics.   The relativistic
Navier-Stokes equations emerge from Einstein's equations   applied to metrics with slowly varying
thermodynamic parameters \refs{\NLFDFG,\BhattacharyyaXC}.  In the context of the AdS/CFT correspondence, this construction also provides an efficient means of computing transport coefficients in the dual CFT.
The details of this connection have been worked out for
pure gravity in arbitrary dimensions \refs{\NLFDFG,\VanRaamsdonkFP, \HaackCP,\BhattacharyyaMZ},  Einstein-Maxwell theory in $4+1$ dimensions \refs{\ErdmengerRM,\BanerjeeTH,\HaackXX}, and
gravity coupled to a scalar field in $4+1$ dimensions  \BhattacharyyaJI.  There is also
an extensive earlier literature on the study of transport coefficients within the AdS/CFT correspondence; see \SonVK\ for a review and further references.

In this paper we consider dyonic black brane solutions in asymptotically AdS$_4$ spacetimes.
These are solutions of Einstein-Maxwell theory in four spacetime dimensions carrying nonzero
charge densities and boundary magnetic fields.
%They were not considered
%in the previous study of charged branes in Einstein-Maxwell theory in arbitrary dimensions because of %the nonzero
%magnetic field.  Indeed, the existence of the dyonic brane solution can be thought of as a consequence
%of electromagnetic duality, which is special to $3+1$ dimensions.
%
The computation of transport coefficients for the dyonic black brane has already received significant
attention in light of its connection to quantum critical points in $2+1$ dimensional condensed matter
systems \refs{\HerzogIJ,\HartnollAI,\HartnollIH,\HartnollIP,\BuchbinderDC}.  The approach taken in
these works differs from ours, and yields results with a different (but overlapping) regime of validity,
as we discuss further below.

Corresponding to the bulk metric and gauge field are a boundary stress tensor and current, which
obey the conservation laws\foot{There is also the dual current, $\tilde{J}^\mu = \epsilon^{\mu \alpha\beta} \nabla_\alpha J_\beta$, but this will play no role in our discussion.}
\eqn\A{\nabla_\mu T^{0\mu} =0~,\quad      \nabla_{\mu} T^{i \mu} = B\epsilon^{ij}J_j~,\quad \nabla_\mu J^\mu =0~,}
where $B$ represents the constant magnetic field on the boundary.   The stress tensor
is also traceless due to scale invariance.
Fluid dynamics is a description of the long wavelength evolution of  conserved charge densities.   At vanishing magnetic field, the conserved charges are energy, momentum, and
electric charge.  Also, by ``long wavelength" one means long compared to the mean free path, which is in
turn proportional to the inverse temperature, $l_{T} \sim {1\over T}$.  One then sets up an expansion
in terms of the small parameter $l_T \p_\mu$, where the derivative acts on the conserved charge densities.    At nonzero magnetic field the story changes.   First, there is now a second length scale in the problem,
$l_B \sim {1\over \sqrt{B}}$.  Second, we see from \A\ that the momentum current $J^{(p_i)\mu} = T^{i\mu}$
is no longer conserved.   At small $B$ there is a separation of the two length scales,   $l_T \ll l_{B}$, and the appropriate fluid dynamical variables depend on the length scale being probed.  At intermediate length
scales $l_T \ll l \ll l_B$, we should treat $ l_T \p_\mu$ as a small parameter but work to all orders in $l_B \p_\mu$.   Also,
in this regime momentum is approximately conserved, and so the momentum density should be kept as a hydroynamical variable; this is the approach used in \refs{\HerzogIJ,\HartnollAI,\HartnollIH,\HartnollIP,\BuchbinderDC}.   At the largest
length scales,  $l\gg l_B$, momentum is no longer an independent hydrodynamical variable, but is instead
fixed in terms of the energy and electric charge densities.  Thus, in this regime we can formulate a simpler
effective theory by ``integrating out" the momentum variables.   This is the approach we take in this paper.
Note that for large magnetic fields, such that $l_B \sim l_T$, there is no separation of scales and it is
clearly inappropriate to include the momenta, since they are non-conserved at leading order.  Keeping the
momentum variables in this case is analogous to retaining a very massive particle in a low energy effective
field theory; it is not  incorrect to do so, but it is  inefficient.  What is being said here is not that
the momentum densities are being set to zero (they are not), but rather that their values are fixed
in terms of the charge and energy densities, and so are not independent.

Our goal is therefore to solve the Einstein-Maxwell equations iteratively in an expansion with respect
to derivatives of the energy and electric charge densities, treating both $l_B \p_\mu$ and $l_T \p_\mu$ as small
expansion parameters.  We carry out this task to second order in the derivative expansion, following the methods applied to
previous examples.  This gives us the stress tensor to second order.  Using the stress tensor
conservation equation, we can then read off the current to third order in derivatives.  This gives us
results for a host of new transport coefficients beyond what was computed in \refs{\HerzogIJ,\HartnollAI,\HartnollIH,\HartnollIP,\BuchbinderDC}, including nonlinear terms.  On the other hand, for the reasons described above, we are restricted to the small frequency/wavelength regime, in contrast
to \refs{\HerzogIJ,\HartnollAI,\HartnollIH,\HartnollIP,\BuchbinderDC} where full AC results can be found at the level of linear response (at least numerically).   Thus, the two  approaches are complementary.

Magnetohydrodynamics is of course a highly developed subject, with many important applications
to astrophysics and plasma physics.   Before trying to compare with this literature it is important to
bear in mind a few points.  First, here we are considering the fluid dynamics of a conformal field theory,
which is quite different than the gas of charged particles considered in most applications.  Second,
most discussion of magnetohydrodynamics are in the context of a dynamical electromagnetic field coupled
to the charged fluid, whereas here we consider a nondynamical external magnetic field.   Note, however, that
that this should be a good model of the dynamical case if the strength of the magnetic field is taken to
be very large while holding the charge fixed, since then we should be able to neglect backreaction from the fluid.

It should be noted that although we usually discuss the case in which only a magnetic field is present
and not an electric field, this is not really a restriction since we can always apply a Lorentz boost
to our solutions to obtain fluid flows with any electric and magnetic fields obeying $E^2 < B^2$.

\newsec{Structure of fluid dynamics}

We begin with a general  discussion of  the structure of our fluid dynamics, independent of the
relation to AdS gravity.  We consider a general $2+1$ dimensional conformal field theory with a
conserved current $J_\mu$ coupled to a constant external magnetic field $B$.  We denote the coordinates of our three dimensional manifold $x^{\mu}$.

\subsec{Derivative expansion}

There are two conserved quantities in our theory,  charge and  energy.  In local thermal equilibrium the corresponding charge and energy densities of our fluid, $\rho$ and $\epsilon$, are slowly varying functions of the coordinates, $x^\mu$.  The current and stress tensor admit an expansion in derivatives of $\eps$ and $\rho$,
and the equations of fluid dynamics are then given by \A.  All that remains to specify the theory are the transport coefficients of the various terms in the derivative expansions of $T^{\mu \nu}$ and $J^{\nu}$.

%We imagine solving the fluid dynamics equations order by order in the derivative expansion, and so
%write
%
%\eqn\am{ \rho = \sum_{n=0}^\infty \rho^{(n)}~,\quad \epsilon = \sum_{n=0}^\infty \epsilon^{(n)}~.}
%
%Our expansion is organized according to the number of $x^\mu$ derivatives plus the sum of the %
%subscripts.

By definition, we have
\eqn\ao{ T^{00} = \epsilon~,\quad J^0 = \rho~,}
with no derivative corrections.\foot{In discussions of fluid dynamics in which the fluid velocity appears as a hydrodynamical variable the charge and energy densities are usually
defined in the fluid rest frame.  In our case there is a preferred  frame, defined such that  there is a magnetic
field but no electric field, and our densities are always measured in this frame.}      To proceed, we need to work out the derivative expansion of
the remaining components,  $T^{0i}$, $T^{ij}$, and $J^i$.

Our task is simplified by noting that  the equations $\p_{\mu} T^{i \mu} = B\epsilon^{ij}J_j$ imply that the spatial components of the current at $n$th
order are related to the stress tensor at order $n-1$,
\eqn\an{ J_i^{(n)}=  -{1\over B}\epsilon_{ij} \left(  \p_0 T^{(n-1)j0}+\p_k T^{(n-1)jk}  \right)~.   }
With this in mind, the remaining equations of fluid dynamics reduce to
\eqn\aza{\eqalign{ &\p_0 \rho + \p_i J^i =0~, \cr & \p_0 \epsilon + \p_i T^{0i} =0~. }}

At zeroth order, tracelessness and rotational symmetry imply
\eqn\ap{ T^{(0)0i} = 0~,\quad T^{(0)ij} ={1\over 2 }\epsilon \delta^{ij}~,\quad J^{(0)i} =0~.}

At first order we have
\eqn\aq{\eqalign{ &T^{(1)0i} = (a_1\delta_{ij}+a_2 \epsilon_{ij})\p_j \epsilon +  (a_3\delta_{ij}+a_4 \epsilon_{ij})\p_j \rho~, \cr & T^{(1)ij}=0 \cr &J^{(1)i} =  -{1\over 2B} \epsilon_{ij}\p_j \epsilon~,  }}
where we used \an. The coefficients $a_{1,2,3,4}$ are functions of $\eps$ and $\rho$ but not their derivatives.

We next consider terms of second order in derivatives.  Here it is convenient to combine the two quantities
$\epsilon$ and $\rho$ into a two component object
\eqn\aqa{q_a = (\epsilon,\rho)~.}
The second order spatial current $J^{(2)i}$ is gotten by applying \an\ to \aq.
Rotational invariance implies that $T^{(2)0i}$ can be built out of the following structures:
\eqn\ar{ \p_0 \p_i q_a~,\quad   \p_0 q_a \p_i q_b~,\quad  \epsilon_{ij} \p_0 \p_j q_a~,\quad   \p_0 q_a\epsilon_{ij}  \p_j q_b~.}
Now, in general, if one is working up to $n$th order, it is permissible to use the zeroth order
equations of motion to simplify the current and stress tensor at $n$th order, since the error
will be of at least order $n+1$.  The zeroth order equations of motion yield $\p_0 \epsilon =\p_0 \rho=0$.
Noting that every term in \ar\ involves at least one time derivative acting on either $\epsilon$ or $\rho$,
we see that we can set $T^{(2)0i}=0$.     This leaves $T^{(2)ij}$, which has the structure
\eqn\as{\eqalign{ T^{(2)ij} &= b_1^{a}[ \p_i \p_j  q_a]^{st} +b_2^a [ \epsilon_{ik}\p_k \p_jq_a]^{st}+ b_3^{ab}[\p_i q_a \p_j q_b]^{st} +b_4^{ab}[ \epsilon_{ik}\p_k q_a \p_j q_b]^{st}~,}}
where we are using the following notation for the symmetric traceless part of a spatial tensor:
\eqn\bm{ [M_{ij}]^{st} = {1\over 2}\left(M_{ij} +M_{ji} - \delta_{ij} M_{kk} \right)~.}

Finally, the third order current is obtained by applying \an\ to the second order stress tensor.

Our fluid dynamics has thus been specified in terms of  the coefficients  $a_{1,2,3,4}$ and
$b_{1,2,3,4}$, which are themselves functions of $\epsilon$ and $\rho$.   In the rest of this
paper we will see that the boundary  stress tensor and current obtained from the Einstein-Maxwell
equations in $4+1$ dimensions falls into the above framework and yields specific formulas for the
coefficients.  In particular, $a_{1,2,3,4}$ are given in equation (6.14); $b_{1,2}$ are given in
equation (6.29); and $b_{3,4}$ are given in equations (6.34) and (6.35).

In the above we considered fluid dynamics in a magnetic field but vanishing electric field.
But by applying a Lorentz boost we can turn on an electric field, subject to the condition
$\vec{E}^2 <B^2$.   As a simple example, consider an infinitesimal Lorentz boost with velocity $v^i$:
$ (t'=t+v\cdot x, x'^i = x^i +v^i t).$ The magnetic field is unchanged, and the new electric field is $E_i = -B\epsilon_{ij}v^j$.
%{\bf (Internal note:  $E_i  = F_{i0}$.)}
To keep things simple, let's
also work to lowest order in both $v^i$ and the derivative expansion (i.e. consider $v^i \sim \p_i \eps \sim \p_i \rho$), in which case we can write the current after the boost as
\eqn\asa{\eqalign{ J_i & = -{1\over 2B} \epsilon_{ij} \p_j \eps -{\rho \over B}\epsilon_{ij} E_j~,}}
which displays the Nernst effect as well as the Hall conductivity.   We can also write down an expression
for the energy flux at this order,
\eqn\asb{T^{(1)0i} = (a_1\delta_{ij}+a_2 \epsilon_{ij})\p_j \epsilon +  (a_3\delta_{ij}+a_4 \epsilon_{ij})\p_j \rho+{3\eps \over B}\epsilon_{ij}E_j~. }
By performing a finite boost and keeping
terms of higher order in the derivative expansion, one can of course easily read off the generalized version of these results, if desired.

We have chosen to formulate our fluid dynamics in terms of the energy and charge density, since these have
a direct physical significance and appear naturally in the gravitational description.  But we can alternatively work in terms of the temperature and chemical potential.  The definition of these is subject to some ambiguity in a non-equilibrium context.  One simple choice is to use the same definition between the energy/charge densities and the temperature/potential as appears in the equilibrium case, without any derivative corrections.  The relation between these sets of variables of course depends on the theory in question.  For the theory corresponding to the dyonic black brane the relations will be given in the following section.

\subsec{Normal modes}

We now look for normal mode solutions by solving  the linearized equations of motion following from \aza.
Dropping nonlinear terms, these become
\eqn\azaa{\eqalign{ &\p_0 \rho +  {a_2 \over B} \vec{\nabla}^2 \p_0 \epsilon + {a_4 \over B} \vec{\nabla}^2\p_0 \rho +{b_2^\eps \over 2B}
(\vec{\nabla}^2)^2  \eps+{b_2^\rho \over 2B}
(\vec{\nabla}^2)^2  \rho=0 \cr  &\p_0 \epsilon +  a_1 \vec{\nabla}^2\eps + a_3 \vec{\nabla}^2\rho=0~.}}
Inserting plane wave fluctuations on top of the constant background,  $\eps+\delta \eps e^{i\vec{k}\cdot \vec{x}-i\omega t}$ and $\rho+\delta \rho e^{i\vec{k}\cdot \vec{x}-i\omega t}$, we find
\eqn\aze{\eqalign{  &(-i\omega{a_2 } -{b_2^\eps \over 2} k^2)k^2 \delta \eps+(iB\omega -ia_4\omega k^2 -{b_2^\rho \over 2} k^4)\delta\rho    =0      \cr &(i\omega + a_1 k^2)\delta\eps + a_3 k^2 \delta\rho =0 ~. }}
Setting the determinant to zero we find two modes
\eqn\azf{\eqalign{ &\omega \approx ia_1k^2 + \cdots~,\quad {\rm with}~~ \delta\rho \approx 0 \cr
&  \omega \approx -{i \over 2} { (a_1 b_2^\rho+a_3 b_2^\eps )\over a_1 B}k^4 +\cdots ~,\quad {\rm with}~~ a_1 \delta\eps + a_3 \delta\rho \approx 0~.}}

In our gravitational computation we will find $a_1 <0$ so that the first mode is strictly decaying in time.
But for the second mode we find that  ${ (a_1 b_2^\rho+a_3 b_2^\eps )\over a_1 B}$ has the same sign as $3(B^2-\rho^2)a^4 +(B^2+\rho^2)(3B^2-\rho^2)$.  Since this can have either sign, the  $k^4$ mode can be
either decaying or growing in time, depending on the values of $\rho$ and $\eps$.

\subsec{Comparison with other approaches}

In this section we have developed a formalism for studying magnetohydrodynamics which, on its surface at least, represents a significant departure from existing treatments in the literature.  While our definitions of the transport coefficients are based on following consistently the logic of hydrodynamic expansions, we would be remiss not to discuss in detail the relation between the different approaches.  The purpose of this section is to provide that discussion, together with a dictionary between the formalisms where appropriate.  We conclude this section with a careful treatment of the shear viscosity and its universal relationship with the entropy density.

In most discussions of relativistic conformal hydrodynamics the starting point is a Lorentz covariant parametrization of the stress tensor,\foot{In this section we are only going to concern ourselves with the stress energy tensor since the conserved current is expressed in terms of the stress energy tensor by equation \an .}  organized in a derivative expansion with respect to the  fluid velocity $u^\mu$ as follows:
\eqn\zaa{\eqalign{
T^{(0)\mu \nu} &= \eps u^{\mu} u^{\nu} + P \Delta^{\mu \nu} \cr
T^{(1)\mu \nu} &= \eta \left( \Delta^{\mu \alpha} \Delta^{\nu \beta}( \p_{\alpha} u_{\beta} + \p_{\beta} u_{\alpha})  - \Delta^{\mu \nu} \p_{\alpha}u^{\alpha} \right),
}}
with
\eqn\zab{
\Delta^{\mu \nu} = \eta^{\mu \nu} + u^{\mu} u^{\nu}.
}
Here, we have listed all of the transport coefficients present to first order in derivatives of the velocity.  A particularly good discussion of the relationship between these expressions and the dynamics of charged fluids can be found in \ErdmengerRM.  We note that the charge density does not enter explicitly until the next order of the derivative expansion.

The form \zaa\ can be justified for fluids in Lorentz invariant backgrounds.  In particular, in thermal
equilibrium the fluid can be taken to have any constant velocity, and the hydrodynamic expansion then
promotes these constants to slowly varying functions. The assumption of a Lorentz invariant background clearly
does not hold in the presence of an external magnetic field; nevertheless, \zaa\ is taken as the starting
point in other treatments of conformal linear magnetohydrodynamics from AdS/CFT \refs{\HartnollIH, \BuchbinderDC,  \BuchbinderNF }.    As we discuss below, it turns out that sensible results can be so obtained
provided that one restricts to the regime of linear fluctuations about thermal equilibrium (which is all
that is considered in \refs{\HartnollIH, \BuchbinderDC,  \BuchbinderNF }), but this approach breaks down
in general, as it does not respect the principles of hydrodynamic expansions.

Focussing first on the shortcomings of \zaa, we note that in the presence of a magnetic field there
is a preferred rest frame, and in thermal equilibrium the fluid will be at rest in this frame.  Since
thermal states are not labeled by an arbitrary constant velocity $u^\mu$, there is no motivation for
introducing a slowly varying velocity field as a hydrodynamical variable, which is the basis of \zaa.
On the other hand, one can consider setting up an expansion in terms of the deviation of the fluid
velocity from its equilibrium value, and this is effectively what is done in
\refs{\HartnollIH, \BuchbinderDC,  \BuchbinderNF }.

Even more problematic than the above is the fact that \zaa\ implies incorrect relationships between transport coefficients due to an inappropriate imposition of Lorentz invariance where there is none.  The presence of an external magnetic field explicitly breaks Lorentz invariance -- there is now a preferred rest frame corresponding to the frame with pure B-field and no E-field.  However, \zaa\ fails to take this into account and is manifestly Lorentz covariant. Explicitly, the vector $v^{\mu} = \eps^{\mu \nu \rho} F_{\nu \rho}$ breaks the Lorentz symmetry, allowing us to separate out the SO(2) vector part $u^i$  of $u^{\mu}$  with a projection operator
\eqn\zac{
\tilde{\Delta}^{\mu \nu} = \eta^{\mu \nu} + {v^{\mu} v^{\nu} \over | v |^2}.
}
With this operator we can separate the transport coefficients according to their SO(2) representations.  For example, if we look for the structure $u^i u^j$ in $T^{ij}$ and $u^0 u^0$ in $T^{00}$, we see that both terms have coefficients fixed by the term $\eps u^{\mu} u^{\nu}$ in \zaa .  However, if we use \zac\ to add additional terms to \zaa\ of the form
\eqn\zab{
\tilde{\Delta}^{\mu}_{\alpha} \tilde{\Delta}^{\nu}_{\beta} u^{\alpha} u^{\beta},
}
then the transport coefficient of $u^i u^j$ in $T^{ij}$ is now independent of the transport coefficient of $u^0 u^0$.

Interestingly,  the imposition of a spurious Lorentz invariance does not cause problems at the level
of linearized fluctuations about thermal equilibrium.  In the preferred frame, we write $u^{\mu} = v^{\mu} + \delta u^{\mu}$, with $\delta u^{0} = 0$ and keep only terms linear in $\delta u^i$.  To lowest order in this expansion $\Delta = \tilde{\Delta}$.  Since the rest frame of the fluid very nearly corresponds with the preferred frame of the B-field, there is no difference between projecting linear fluctuations (or derivatives of fluctuations) to either the rest frame or the preferred frame since any difference will be second order.  At any order in derivatives, SO(2) invariance lifts to lorentz invariance for linear fluctuations.  This phenomena explains why other authors have not run into any trouble using \zaa : they were considering only linearized fluctuations.  In order to study manifestly nonlinear properties though, we needed a new formalism. In particular, we avoid the problems described above by imposing only the actual $SO(2)$ symmetry, and using
as hydrodynamical variables only those quantities that can be freely specified in equilibrium, namely the
energy and charge density.

With the preceding comments in mind, it is now interesting to compare results obtained by these different
approaches.
To bridge the gap between the two formalisms, we define the fluid velocity in our formalism by equating the velocity with the Lorentz boost parameter needed to annihilate $T^{0i}$.  This definition is equivalent to matching \zaa\ with \aq\ and reading off
\eqn\zac{
u^i = {1 \over \eps + P} T^{0i} = {2 \over 3 \eps} \left[(a_1\delta_{ij}+a_2 \epsilon_{ij})\p_j \epsilon +  (a_3\delta_{ij}+a_4 \epsilon_{ij})\p_j \rho \right].
}
This expression is of course corrected at higher orders in the derivative expansion, but this version is sufficient for our current purposes.  It is worth noting that this velocity is manifestly small in the hydrodynamic expansion regardless of the amplitude of any fluctuations as it is proportional to the derivatives of our charge and energy densities.

With our definition of the velocity in hand, we can now use \zaa\ to find a suitable definition for the shear viscosity.  Since \zaa\ is valid for linear fluctuations, we can use \zac\ as a dictionary to relate the shear viscosity $\eta$ to our linear transport coefficients.  If we plug \zac\ into \zaa\ and keep only terms linear in our fluctuations then we have
\eqn\zad{
T^{(2) ij} = {4 \eta \over 3}\left[(a_1\delta_{ik} + a_2 \epsilon_{ik})\p_i \p_k \epsilon +  (a_3\delta_{ik}+a_4 \epsilon_{ik})\p_i \p_k \rho \right]^{st}.
}
Note that a first order contribution in a hydrodynamic expansion with respect to the velocities is equivalent here to a second order contribution with respect to the charge and energy densities since the velocity itself is now being considered first order in the hydrodynamic expansion.  As noted above, second order linear contributions to the stress tensor from derivatives of the charge density were not included in \zaa , so the second half of \zad\ is not predictive of any relationship with our current results.  This means that we should only interpret the shear viscosity as a factor in the diffusion of energy and not charge.  Since the only terms contributing to the second order linear energy transport coefficients from the standard formalism are those in \zad, we can read off relationships between our transport coefficients $b_1^{\eps}, b_2^{\eps}$ and the shear viscosity $\eta$.  Comparison with \as\ yields
\eqn\zae{
b_1^{\eps} = {4 \eta a_1 \over 3} ~, \quad b_2^{\eps} = {4 \eta a_2 \over 3}~.
}
If we skip ahead  and insert the values for these transport coefficients that we obtain from AdS/CFT in section 6, we see that both of these equations imply that
\eqn\zae{
\eta = {a^2 \over 4}~,
}
where $a$ is related to the entropy density $s$ according to $s = \pi a^2$.  Putting these facts together yields the universal ratio
\eqn\zaf{
{\eta \over s} = {1 \over 4 \pi}~.
}
This result is a slight extension of  \BuchbinderNF\ which demonstrated this result for uncharged fluids in the presence of a magnetic field.  It should be noted however that while our result takes into account additional degrees of freedom in the form of a variable charge density, we have not used any nonlinear effects to demonstrate this relationship.  As noted above, the shear viscosity is a purely linear property of our system and has no dependence on nonlinear properties.

\newsec{Gravity side}

We now turn to the gravitational description of fluid dynamics.  In this section we write down the
dyonic black brane solution and its thermodynamic properties, and give the general prescription for computing
the boundary stress tensor and current.

\subsec{Action and dyonic black brane solution}

The bulk Maxwell-Einstein action with a negative cosmological constant, which can be thought of as a
consistent truncation of the theory resulting from an $S^7$ compactification of M-theory,    is
\eqn\aa{
 S= {2 \over \kappa_4^2} \int d^4x \sqrt{-g} \left[{1 \over 4} R - {1 \over 4} F_{M N} F^{M N} - {3 \over 2L^2} \right]~.
}
We henceforth choose units with $L=1$.
The equations of motion are
\eqn\ab{\eqalign{
W_{MN}&\equiv R_{MN} + 3 g_{MN} - 2 F_{M P} F_{N}^{~P} + \half g_{MN} F_{PQ}F^{PQ}=0 \cr
 Y^N&\equiv \nabla_M F^{MN} = 0~.}}

The dyonic black brane solution in Eddington-Finkelstein coordinates is
\eqn\ac{\eqalign{
ds^2 &= 2 dv dr - U(r) dv^2 + r^2dx^i dx^i \cr F & = {\rho \over r^2} dr \wedge dv +B dx^1 \wedge dx^2
}}
where U(r) is given by
\eqn\aca{
U(r)= r^2 + {\rho^2 + B^2 \over r^2} - {2\epsilon \over  r}~.
}
The horizon is located at the largest real root of the equation $U(r)=0$.  Calling this $r=a$, $a$ is thus given
by the largest real root of
\eqn\ad{  a^2 + {\rho^2 + B^2 \over a^2} - {2\epsilon \over  a}=0~.}
The energy and charge densities (defined below) are given by $\epsilon$ and  $\rho$.

The Hawking temperature is
\eqn\ae{
T = {3 a \over 4 \pi} - {B^2 + \rho^2 \over 4 \pi a^3}~.
}
$\eps$ and $\rho$ are restricted to values such that $T\geq 0$.
The chemical potential can be read off from the asymptotic value of $A_v$ in a gauge such that $A_v$
vanishes at the horizon.  This gives $\mu = {\rho \over a}$.

Some further conventions:
Latin  indices $M,N,...$ run over all four spacetime  coordinates, while Greek indices $\mu, \nu, ...$
run over the three coordinates $(v, x^1,x^2)$.   Since $v$ plays the role of time on the boundary,
we will sometimes use  $v=x^0$.   The boundary theory will always see a Minkowski metric,
$\tilde{\gamma}_{\mu\nu}dx^\mu dx^\nu = -(dx^0)^2+dx^i dx^i$.    Indices on the boundary stress
tensor and currents are raised and lowered with this metric.

\subsec{Stress tensor and current}

The action \aa\ should be supplemented with the boundary terms
\eqn\af{
S_{bndy} =  - {1 \over \kappa_4^2} \int_{\p M} d^3 x \sqrt{-\gamma} \theta - {2 \over \kappa_4^2} \int_{\p M} d^3 x \sqrt{-\gamma}~.
}
Here $\gamma$ is the boundary metric and $\theta = \gamma^{\mu\nu}\theta_{\mu\nu}$, where
$\theta_{\mu\nu}= - {1\over 2} ( \del_{\mu} n_{\nu} + \del_{\nu} n_{\mu})$ is the extrinsic curvature of the boundary, defined in terms of the outward pointing unit normal vector $n$.

The conformal boundary metric is defined as $\tilde{\gamma}_{\mu\nu} = \lim_{r\rightarrow \infty} {1\over r^2} \gamma_{\mu\nu}$.  Also, the boundary gauge field is defined as $\lim_{r\rightarrow \infty}A_\mu$, in a gauge where $n^M A_M=0$.    The boundary stress tensor and current are then defined as
\eqn\ag{
\delta S = {1 \over \kappa_4^2} \int_{\p M} \sqrt{-\tilde{\gamma}} \left(2 J^{\mu} \delta A_{\mu} +
T^{\mu \nu} \delta \tilde{\gamma}_{\mu \nu}\right)~.
}
Explicitly \BalasubramanianRE,
\eqn\ah{\eqalign{
J^{\mu} &= r^2 F^{\mu r} \cr
T^{\mu \nu} &= {r^5 \over 2} \left[ \theta^{\mu \nu} - \theta \gamma^{\mu \nu} - 2 \gamma^{\mu \nu} \right]~.
}}
Implicit in \ah\ is the large $r$ limit, as well as a projection of $T^{\mu\nu}$ parallel to the boundary (since the orthogonal component does not appear in \ag.)

Electromagnetic gauge invariance implies current conservation,
\eqn\ai{ \nabla_\mu J^\mu =0~.}
Invariance under diffeomorphisms generated by vector fields tangent to the boundary yields the (non) conservation equation
\eqn\aj{ \nabla_{\nu} T^{\mu \nu} = F^{\mu \nu} J_{\nu}~.}
Tracelessness of the stress tensor follows from invariance under diffeomorphisms shifting the radial location of the boundary
\eqn\ak{ \tilde{\gamma}_{\mu\nu} T^{\mu\nu} =0~.}
In particular, the latter invariance follows from the absence of logarithmic divergences in the bulk action, the presence of which would necessitate adding a non-diff invariant counterterm \HenningsonGX.

Applied to the solution \ac\ we find
\eqn\al{\eqalign{ T^{\mu\nu} &= {\rm diag} (\epsilon, {1\over 2} \epsilon, {1\over 2}\epsilon) \cr J^\mu &= (\rho , 0 ,0)~.}}

\newsec{Gravitational derivative expansion}

The dyonic black brane solution above corresponds to a fluid in thermal equilibrium.  To find a solution corresponding to local thermal equilibrium, we follow the logic in \NLFDFG\ and construct a solution which looks locally like the dyonic black brane.  We take the expression in \ac\ and allow $\eps$ and $\rho$ to be functions of the boundary coordinates $x^{\mu}$.  While this expression is no longer a solution of the Maxwell-Einstein equations, it is approximately a solution in the limit that derivatives of $\eps$ and $\rho$ are small.  By adding small corrections order by order in a derivative expansion we construct a perturbative solution which locally approximates the dyonic black hole.

To implement the derivative expansion, we formally regard $\epsilon$ and $\rho$ as being functions of $\ve x^\mu$, where $\ve$ is a formal expansion parameter that is eventually set equal to $1$.   The metric and
gauge fields are expanded as
\eqn\at{\eqalign{ g & =g^{(0)}(\eps,\rho) + \ve g^{(1)}(\eps,\rho)+ \ve^2 g^{(2)}(\eps,\rho) + {\cal O}(\ve^3) \cr A & = A^{(0)}(\eps,\rho) + \ve A^{(1)}(\eps,\rho)+ \ve^2 A^{(2)}(\eps,\rho) + {\cal O}(\ve^3)~,}}
where $g^{(0)}$ and  $A^{(0)}$  represent the lowest order solution given in \ac.  The energy and charge densities are themselves given by an expansion,
\eqn\au{ \epsilon = \eps^{(0)}(\ve x^\mu)+ \ve \eps^{(1)}(\ve x^\mu) + \cdots ~,\quad\rho = \rho^{(0)}(\ve x^\mu)+ \ve \rho^{(1)}(\ve x^\mu) + \cdots }
As explained in \NLFDFG, the equations of motion can be solved ``tubewise" by working near
a given $x^\mu$ location, say $x^\mu=0$.  It is then convenient to set $\eps^{(n>0)}(0)= \rho^{(n>0)}(0)=0$.
In the following, the order of a quantity is defined with respect to its associated power of $\ve$. Note that this is slightly different than the labelling used in section 2 where we just counted the number of derivatives
acting on $\eps$ and $\rho$.

For the metric and gauge fields we make the gauge choice
\eqn\av{ A_r =0~,\quad g_{rr}=0~,\quad {g^{(0)}}^{\mu\nu}g_{\mu\nu}^{(n>0)} =0 ~.}

The zeroth order solution preserves $SO(2)$ rotational symmetry, and this can be used to classify
the corrections to the metric and gauge fields.   The gauge field corrections are written
\eqn\aw{ A^{(n)} = A_v^{(n)} dv+ A_i^{(n)} dx^i~,}
with $A_v^{(n)}$  an $SO(2)$ scalar and $A_i^{(n)}$ an $SO(2)$ vector.     For the metric we write
\eqn\ax{(ds^2)^{(n)}={k^{(n)}\over r^2}dv^2-2h^{(n)}dvdr +r^2 h^{(n)} dx^i dx^i + 2j^{(n)}_idv dx^i
+r^2 \sigma_{ij}^{(n)} dx^i dx^j~.}
In this expansion, $k^{(n)}$ and $h^{(n)}$ are $SO(2)$ scalars;  $j^{(n)}_i$ is an $SO(2)$ vector; and
$\sigma_{ij}^{(n)}$ is an $SO(2)$ symmetric traceless tensor.

We impose the following large $r$ boundary conditions on the $n>0$ components:
\eqn\ay{ \eqalign{ &A_v^{(n)} \sim {1\over r^2}~,\quad A_i^{(n)} \sim {1\over r}  \cr & k^{(n)} \sim r^0~ \quad h^{(n)} \sim {1\over r^4}~,\quad j^{(n)}_i  \sim {1\over r}~,\quad \sigma^{(n)}_{ij} \sim {1\over r^3}~.}}
These conditions follow from a combination of the asymptotic AdS boundary conditions along with
the freedom to redefine coordinates as well as the zeroth order solution, as in \VanRaamsdonkFP.

%\HorowitzJD
\lref\HorowitzJD{
  G.~T.~Horowitz and V.~E.~Hubeny,
  ``Quasinormal modes of AdS black holes and the approach to thermal
  equilibrium,''
  Phys.\ Rev.\  D {\bf 62}, 024027 (2000)
  [arXiv:hep-th/9909056].
  %%CITATION = PHRVA,D62,024027;%%
}

In addition to these large $r$ boundary conditions, we must also demand that our solution be smooth across the horizon at $r=a$ in order to uniquely determine a solution.  As shown in \HorowitzJD, this condition is equivalent to demanding the presence of purely ingoing modes at the future horizon.

%\WittenYA
\lref\WittenYA{
  E.~Witten,
  ``SL(2,Z) action on three-dimensional conformal field theories with Abelian
  symmetry,''
  arXiv:hep-th/0307041.
  %%CITATION = HEP-TH/0307041;%%
}

Although the Maxwell-Einstein equations of motion  are invariant under electric-magnetic duality, the
boundary conditions \ay\ are not.   In particular, our boundary conditions fix a constant
magnetic field and vanishing electric field on the boundary.  However, electric-magnetic duality
will turn on a nonzero electric field in general.  See \WittenYA\ for further discussion of the breaking
of duality by AdS/CFT boundary conditions.

We then find  the current and stress tensor (a large $r$ limit is implicit):
\eqn\az{\eqalign{J^{v}& =\rho
%r^2 \p_r A^{(n)}_v
\cr J^{i}& = -r^2\sum_n \p_r A^{(n)}_i \cr T^{vv} &=\eps
\cr T^{vi} &= -{3\over 4} r \sum_n j^{(n)}_i \cr T^{ij} &={1\over 2}\eps+
{3\over 4} r^3 \sum_n \sigma^{(n)}_{ij}   }}
%
%The vanishing of $J^{(n)v}$ and $T^{(n)vv}$ for $n>0$ is a consequence of our definitions
%$J^v = \rho$ and $T^{vv}=\eps$ along with the conditions $ \rho^{(n>0)}(0)=\eps^{(n>0)}(0)=0$.

\newsec{Structure of perturbation theory}

The $n$th order metric coefficients are determined by the components of the Einstein equations $W^{(n)}_{MN}=0$ with $M, N \neq v$.         These equations can be organized as
\eqn\ba{\eqalign{ W^{(n)}_{rr}& = -{1\over r^4}\p_r(r^4 \p_r h^{(n)}) - S^{(n)}_{(h)} =0
\cr
r^2 (UW_{rr})^{(n)}-W_{ii}^{(n)}& = \p_r \left(-{2\over r}k^{(n)}\right)  +\p_r \left(\p_r(r^2 U^{(0)}) h^{(n)}\right)-{8\over r^2}B^2 h^{(n)} +  4\rho^{(0)}\p_r A_v^{(n)}  - S^{(n)}_{(k)} =0
\cr
W^{(n)}_{ri}& ={1\over 2}r \p_r \left({1\over r^2}\p_r (r j_i^{(n)})\right) +{2\over r^2 } [\rho^{(0)}\delta_{ij}-B\epsilon_{ij}]\p_r A_j^{(n)}  - S^{(n)}_i =0
\cr
W^{(n)}_{ij}-{1\over 2}\delta_{ij} W^{(n)}_{kk}& =\p_r \left(-{1\over 2}r^2 U^{(0)}\p_r \sigma_{ij}^{(n)}\right) - S^{(n)}_{ij} =0
}}
The source terms appearing above are constructed from the solution at order $n-1$, and so are assumed
to be known.

Similarly,  two components of the Maxwell equations yield
\eqn\bb{\eqalign{ Y^{(n)v} &= {1\over r^2} \p_r \left(-r^2 \p_r A_v^{(n)}-2\rho^{(0)}h^{(n)}\right) - V^{(n)}=0 \cr Y^{(n)i}&= {1\over r^2} \p_r\left( U^{(0)} \p_r A_i^{(n)} +{1\over r^2}[\rho^{(0)}\delta_{ij}+B\epsilon_{ij}]j^{(n)}_j  \right) - V^{(n)}_i=0   }}
These equations, together with the boundary conditions \ay, are sufficient information to solve
for all the metric and gauge field functions in terms of $\eps$ and $\rho$.    The remaining
Einstein-Maxwell equations then becomes conditions on $\eps$ and $\rho$; in particular,  they are identified with  equations of fluid dynamics \A\ expanded to a given order in $\ve$.

\newsec{Solving the equations}

\subsec{Zeroth order solution}

At zeroth order we use the solution \ac\ but with $\eps = \eps^{(0)}$ and $\rho= \rho^{(0)}$.  By construction, this is a solution to the Einstein-Maxwell equations.  The current and stress tensor are
\eqn\bbb{\eqalign{ J^{\mu} &= (\rho,0,0) \cr  T^{\mu\nu}&= {\rm diag}(\eps,{1\over 2} \eps,{1\over 2} \eps)~.}}

\subsec{First order solution}

At first order we write
\eqn\bba{\eqalign{ \eps(x^\mu)&=  \eps^{(0)} + \ve x^\mu \p_\mu \eps^{(0)}(0) \cr \rho(x^\mu)& = \rho^{(0)} + \ve x^\mu \p_\mu \rho^{(0)}(0) ~.}}
%
%Further, we can always set $\eps^{(1)}(0)=\rho^{(1)}(0)=0$ by absorbing these into the zeroth order
%quantities.

The first order sources are built out of $\p_\mu \eps^{(0)}$ and $\p_\mu \rho^{(0)}$, and read
\eqn\bc{\eqalign{S^{(1)}_{(h)}& = S^{(1)}_{(k)} =  S^{(1)}_{i} =   S^{(1)}_{ij} =V^{(1)} = 0~,\cr   V^{(1)}_i& ={\p_i \rho^{(0)}\over r^4}                }}
All of the functions in \ba-\bb\ appearing without sources can be set to zero, either due to the boundary
conditions or by absorbing into the lowest order solution.    What remains is  to find
$\p_r A^{(1)}_i$ and $j^{(1)}_i$.

Integrating $Y^{(1)i}=0$ gives
\eqn\bd{ U^{(0)} \p_r A_i^{(1)} +{1\over r^2}[\rho^{(0)}\delta_{ij}+B\epsilon_{ij}]j^{(1)}_j +{\p_i \rho^{(0)}\over r} = -c^{(1)}_i~,}
where  $c_i$ is $r$-independent.   Solving for $\p_r A_i^{(1)}$ and plugging into $W^{(1)}_{ri}=0$ gives
\eqn\be{ U^{(0)}\p_r^2 j_i^{(1)} - U''^{(0)} j^{(1)}_i = {4\over r^2}(\rho^{(0)}\delta_{ij}-B\epsilon_{ij})(c_j^{(1)} +{1\over r}\p_j \rho^{(0)}) }
where  $'$ denotes the radial derivative and we used the following identity for $U$:
\eqn\bff{ U''  = {2 U \over r^2} +{4\over r^2}(\rho^2 +B^2)~.}

The general solution of \be\ is
\eqn\bg{ j^{(1)}_i(r) = -U^{(0)}(r) (\rho^{(0)}\delta_{ij}-B\epsilon_{ij}) \int_{\alpha^{(1)}_j}^r \! dr'{\beta^{(1)}_j +{4\over r'}c_j^{(1)} +{2\over r'^2} \p_j \rho^{(0)} \over U^{(0)}(r')^2}~,}
with integration constants $\alpha^{(1)}_j$ and $\beta^{(1)}_j$.  The integral is elementary but lengthy ---
see the appendix.
The large $r$ boundary condition
fixes $\alpha^{(1)}_j=\infty$.   $\beta^{(1)}_j$ is fixed by demanding regularity at the horizon.
In particular, for generic $\beta^{(1)}_j$   we find that $j^{(1)}_i$ has a logarithmic term,
$j^{(1)}_i \sim (r-a) \ln(r-a)$, which yields a divergent derivative.   We can cancel this term
by choosing
\eqn\bh{ \beta^{(1)}_i = -{4 \over a^{(0)}}\left(1+ {U'^{(0)}(a^{(0)}) \over a^{(0)}U''^{(0)}(a^{(0)})}\right)c_i^{(1)}-{2 \over {a^{(0)}}^2}\left(1+ {2U'^{(0)}(a^{(0)}) \over a^{(0)}U''^{(0)}(a^{(0)})}\right)\p_i \rho^{(0)}~. }
With this choice of integration constants we find that  $j^{(1)}_i$ has the large $r$ behavior
\eqn\bi{ j^{(1)}_i(r)  = {(\rho^{(0)}\delta_{ij}-B\epsilon_{ij}) \beta^{(1)}_j \over 3r} +{\cal O}({1\over r^2})~.}
$\p_r A^{(1)}_i$ can now be found from \bd.

Up to first  order the current and stress tensor can now be computed as
\eqn\bia{\eqalign{  &J^{i} =c^{(1)}_i      \cr &T^{vi}=-{1\over 4}(\rho^{(0)}\delta_{ij}-B\epsilon_{ij}) \beta^{(1)}_j   \cr &[T^{ij}]^{st}=0~. }}

The remaining Maxwell-Einstein equations are $W^{(1)}_{vM}=0$ and $Y^{(1)r}=0$.  With the above
expressions  for $j^{(1)}_i$ and  $\p_r A^{(1)}_i$ we can verify that these  equations are
equivalent to the  order $\ve$ part of the  equations
\eqn\bj{ \p_\mu J^{\mu}=0~,\quad \p_\nu T^{\mu\nu} = F^{\mu\nu} J_\nu~.}
%
%where $F^{\mu\nu}$ corresponds to the boundary magnetic field, and the  current and stress tensor are given by %their lowest order expressions \bbb.
Note that for the powers of $\ve$ to match up, the  stress tensor and current appearing with a derivative are given by their lowest order expressions  \bbb, while the undifferentiated current is given by \bia.   In particular, these equations imply
\eqn\bk{\eqalign{ &\p_v \rho^{(0)} =\p_v \eps^{(0)} = 0 \cr &c^{(1)}_i  = -{1\over 2B}\epsilon_{ij} \p_j \eps^{(0)}~. }}

We can now read off the first order transport coefficients by writing
\eqn\bka{T^{vi} = (a_1\delta_{ij}+a_2 \epsilon_{ij})\p_j \epsilon +  (a_3\delta_{ij}+a_4 \epsilon_{ij})\p_j \rho~, }
as in \aq.   Using the explicit form of $U$ as well as the relation between  $\eps$ and $a$, we find
\eqn\bkb{\eqalign{a_1 &= -{3\over 4}\left({\eps \over \rho^2 +B^2}\right) \cr a_2 &= -{3\over 4} \left({\eps \over \rho^2 +B^2}\right){\rho \over B}\cr a_3 & = {1\over 2} \left({\rho \over \rho^2 +B^2}\right)\left({ 3a \eps -\rho^2 -B^2\over a^2}\right) \cr
a_4 & = -{1\over 2} \left({B \over \rho^2 +B^2}\right)\left({3a\eps -\rho^2 -B^2\over a^2}\right)~.}}

\subsec{Second order solution}

$\eps$ and $\rho$ are now given by expanding out \au\ to order $\ve^2$,
\eqn\bkd{  \eqalign{ \eps(x^\mu)&=  \eps^{(0)}(0) + \ve x^\mu \p_\mu \eps^{(0)}(0)+ {1\over 2}\ve^2x^\mu x^\nu \p_\mu \p_\nu\eps^{(0)}(0)+ \ve^2 x^\mu \p_\mu \eps^{(1)}(0)  \cr \rho(x^\mu)& = \rho^{(0)}(0) + \ve x^\mu \p_\mu \rho^{(0)}(0) + {1\over 2}\ve^2x^\mu x^\nu \p_\mu \p_\nu\rho^{(0)}(0)+ \ve^2 x^\mu \p_\mu \rho^{(1)}(0) ~.}}

The second order sources work out to be
\eqn\bl{\eqalign{S^{(2)}_{(h)}& = {2\over r^2}(\p_r A_i^{(1)})^2 \cr
S^{(2)}_{(k)}& =2U^{(0)}(\p_r A^{(1)}_i)^2 + {4 \over r^2} \rho^{(0)}\p_r A^{(1)}_i j^{(1)}_i-{4 \over r^2}B\epsilon_{ij}\p_i A^{(1)}_j \cr &\quad  +{2\over r} \p_i j^{(1)}_i  +\p_r \p_i j^{(1)}_i  -{2\over r} j^{(1)}_i \p_r j^{(1)}_i -{1\over 2}(\p_r j^{(1)}_i)^2 \cr
S^{(2)}_{i}& = 0\cr
S^{(2)}_{ij}& = -\p_r [\p_i j^{(1)}_j]^{st} -{2\over r} [j^{(1)}_i \p_r j^{(1)}_j]^{st} +{2\over r^2} [j^{(1)}_i j^{(1)}_j]^{st}+{1\over 2}[\p_r j^{(1)}_i \p_r j^{(1)}_j]^{st} \cr &+2U^{(0)}[\p_r A^{(1)}_i \p_r A^{(1)}_j]^{st}+{4\over r}[\p_i \rho^{(0)} \p_r A^{(1)}_j]^{st} +{4\over r^2} B[\p_r A^{(1)}_i \epsilon_{jk}j^{(1)}_k]^{st} \cr
V^{(2)}& ={1\over r^2}\p_r \p_i A_i^{(1)}-{1\over r^2}\p_r(j^{(1)}_i \p_r A_i^{(1)}) \cr
V^{(2)}_i& =   0~.         }}

We can solve the equations separately in the scalar, vector, and symmetric traceless tensor sectors.

\vskip.2cm
\noindent ${\bf {\underline {scalar~ sector:}}}$
\vskip.2cm

We determine $h^{(2)}$ from the equation $W^{(2)}_{rr}=0$, which takes the form
\eqn\bn{ \p_r (r^4 \p_r h_2) =-2r^2 (\p_r A^{(1)}_i)^2~.}
The source on the right hand side is determined from \bd, \bg, and \bk.   Of main relevance is the
fact that $\p_r A^{(1)}_i$ is a smooth function, including at the horizon, and has asymptotic behavior
\eqn\bo{ \p_r A^{(1)}_i = -{c_i^{(1)}\over r^2} + {\cal O}({1\over r^3})~.}
We can therefore proceed by straightforward integration:
\eqn\bp{ h^{(2)}(r) = -2 \int_\infty^r \! {dr' \over  r'^4} \int_\infty^{r'} \! dr'' \left( r'' \p_r A_i^{(1)}(r'')\right)^2~.}

Next, $\p_r A_v^{(2)}$ is found from integrating $Y^{(2)v}=0$.   Writing this equation
as ${1\over r^2} \p_r \left(-r^2 \p_r A_v^{(2)}\right) + X_1 =0$, the solution with the
required boundary conditions is
\eqn\bpa{ \p_r A_v^{(2)}(r) = {1\over r^2} \int_\infty^r \! dr' r'^2 X_1(r')~.}

Finally, $k^{(2)}$ is found by integrating $r^2 (UW_{rr})^{(2)}-W_{ii}^{(2)}=0$.  Writing this equation in the form $\p_r \left(-{2\over r}k^{(2)}\right) + X_2 =0$ we can straightforwardly integrate to get
\eqn\bq{ k^{(2)}(r) = {1\over 2}r \int_\infty^r \! dr' X_2(r')~.}

This completes the solution in the scalar sector.  The explicit integrals appearing above are complicated and not especially illuminating.  Since these functions do not show up
in the stress tensor and current it is not necessary to compute them explicitly in order to
find the second order transport coefficients.

\vskip.2cm
\noindent ${\bf {\underline {vector~ sector:}}}$
\vskip.2cm

The vector sector is simple since the corresponding source terms in \bl\ are zero.  The solution
is obtained as in the first order computation, with the only difference that we do not have
the analog of the source terms proportional to $\p_i \rho^{(0)}$.   We thus have
\eqn\bqa{ j^{(2)}_i(r) = -U^{(0)}(r) (\rho^{(0)}\delta_{ij}-B\epsilon_{ij}) \int_{\infty}^r \! dr'{\beta^{(2)}_j +{4\over r'}c_j^{(2)}\over U^{(0)}(r')^2}~,}
with
\eqn\bqb{ \beta^{(2)}_i = -{4 \over a^{(0)}}\left(1+ {U'^{(0)}(a^{(0)}) \over a^{(0)}U''^{(0)}(a^{(0)})}\right)c_i^{(2)}~. }
$\p_r A_i^{(2)}$ is determined by the  analog of \bd,
\eqn\bqc{ U^{(0)} \p_r A_i^{(2)} +{1\over r^2}[\rho^{(0)}\delta_{ij}+B\epsilon_{ij}]j^{(2)}_j = -c^{(2)}_i~.}
$\beta^{(2)}_i$ and $c_i^{(2)}$ now contribute to the stress tensor and current precisely as in \bia.
Also, the stress tensor conservation equation now fixes
\eqn\bqe{ c_i^{(2)} = -{1\over 2B} \epsilon_{ij} \p_j \eps^{(1)}~.}
just as in \bk.  Therefore, the vector components of the stress tensor and current at this order take the
same form as at first order,
\eqn\bqd{\eqalign{  &J^{i} =-{1\over 2B} \epsilon_{ij} \p_j \eps     \cr &T^{vi} = (a_1\delta_{ij}+a_2 \epsilon_{ij})\p_j \epsilon +  (a_3\delta_{ij}+a_4 \epsilon_{ij})\p_j \rho~~. }}
with the same coefficients \bkb.  In writing this expression for $T^{vi}$ we have used
$\rho^{(1)}(x^\mu)=0$.  This is consistent, since the equation $\p_\mu J^\mu=0$ at order $\ve^2$ is
simply $\p_v \rho^{(1)}=0$, as follows from the fact that the spatial current is divergenceless at this order.

This completes the solution in the vector sector.

\vskip.2cm
\noindent ${\bf {\underline {tensor~ sector:}}}$
\vskip.2cm

The tensor components are determined by solving $W^{(2)}_{ij}-{1\over 2} \delta_{ij}W_{kk}^{(2)}=0$.
Integrating, we obtain
\eqn\br{\sigma_{ij}^{(2)}(r) = -2 \int_\infty^r  {dr' \over r'^2 U^{(0)}(r')} \int_{a^{(0)}}^{r'} \! dr'' S^{(2)}_{ij}(r'')~.}
The lower integration limit on $r''$ is chosen to avoid a singularity at the horizon, while the
integration limit for $r'$ is chosen to impose the correct asymptotic behavior.

Of primary interest is the contribution to $T^{ij}$ given in \az.   This gives
\eqn\bs{ [T^{ij}]^{st} ={1\over 2}\int_{a^{(0)}}^{\infty}\! dr ~S^{(2)}_{ij}(r)~.  }
From this we can read off transport coefficients $b_{1,2,3,4}$ appearing in \as.

The simplest contributions are the second derivative terms,  $b_1^a [\p_i \p_j q_a]$  and $b_2^a [ \epsilon_{ik}\p_k \p_jq_a]^{st}$,  which only get a contribution from the source term $S^{(2)}_{ij} = -\p_r [\p_i j^{(1)}_j]^{st}$.  We find
\eqn\bt{\eqalign{b_1^\eps & = {1 \over {a^{(0)}}^2 U''^{(0)}(a^{(0)}) } ={1\over 4}{a^2 \over (\rho^2+B^2)}  \cr
b_1^\rho & = -{2 \rho^{(0)} \over {a^{(0)}}^3 U''^{(0)}(a^{(0)}) }=-{1\over 2}{a \rho\over (\rho^2+B^2)} \cr
b_2^\eps & = { \rho^{(0)} \over {a^{(0)}}^2 U''^{(0)}(a^{(0)}) B} ={1\over 4}{a^2 \over (\rho^2+B^2)} {\rho \over B}\cr
b_2^\rho & = {2 B \over {a^{(0)}}^3 U''^{(0)}(a^{(0)}) }={1\over 2}{a B\over (\rho^2+B^2)}~.  }}

The remaining nonlinear transport coefficients are much more tedious to obtain.  There are
two contributions to the stress tensor: $[T^{ij}]^{st}= [T_1^{ij}]^{st}+[T_2^{ij}]^{st}$.

 The contribution
from the source term $S^{(2)}_{ij} = -\p_r [\p_i j^{(1)}_j]^{st}$ is easy to obtain, and yields the
stress tensor
\eqn\bu{\eqalign{ [T_1^{ij}]^{st}& = {1\over 4} {\p \over \p \eps}\left({a(\eps,\rho)^2 \over \rho^2+B^2}\right) [\p_i\eps  \p_j \eps ]^{st}
 -{1\over 2} {\p \over \p \rho}\left({a(\eps,\rho) \rho \over \rho^2+B^2}\right) [\p_i\rho  \p_j \rho ]^{st}
\cr &+\left[ {1\over 4} {\p \over \p \rho}\left({a(\eps,\rho)^2 \over \rho^2+B^2}\right) -{1\over 2}  {\p \over \p \eps}\left({a(\eps,\rho)\rho \over \rho^2+B^2}\right)  \right][\p_i \eps \p_j \rho]^{st}
\cr &
+{1 \over 4B}{\p \over \p \eps}\left({a(\eps,\rho)^2 \rho  \over \rho^2+B^2}\right)  [\epsilon_{ik}\p_k \eps \p_j \eps ]^{st}
+{B \over 2}{\p \over \p \rho}\left({a(\eps,\rho) \over \rho^2+B^2}\right)  [\epsilon_{ik}\p_k \rho\p_j \rho ]^{st} \cr
& + \left[{1 \over 4B}{\p \over \p \rho}\left({a(\eps,\rho)^2 \rho  \over \rho^2+B^2}\right) +{B \over 2}{\p \over \p \eps}\left({a(\eps,\rho) \over \rho^2+B^2}\right)  \right][\epsilon_{ik}\p_k \eps \p_j \rho]^{st}~.
}}
Here $a(\eps,\rho)$ is obtained by solving \ad.

The rest of the contribution $T_2^{ij}$  take the form of integrals, which we will not evaluate explicitly.  The sources are built up out of $j_i^{(1)}$ and $\p_r A_i^{(1)}$.  These were found in
the first order analysis, and we write them in
the form
\eqn\bv{\eqalign{ j_i^{(1)} & =  (\rho \delta_{ij}-B \epsilon_{ij})(j_{\epsilon}(r) \eps_{j k} \p_k \eps + j_\rho(r)\p_j \rho ) \cr     \p_r A^{(1)}_i & = A_{\eps}(r) \eps_{ij} \p_j \eps + A_\rho(r) \p_i\rho  }}
with
\eqn\bw{\eqalign{ j_{\eps}(r)& = {2 \over B a} U(r) \int_\infty^r \! dr' \left({ 1  +{ U'(a)\over a U''(a)} -{a \over r'}  \over U(r')^2} \right)    \cr
j_\rho(r)& = {2\over a^2} U(r) \int_\infty^r \! dr' \left({ 1  +{2 U'(a)\over a U''(a)} -{a^2 \over r'^2}  \over U(r')^2} \right) \cr
 A_{\eps}(r)& = - {1\over 2 B U(r)} -{1\over 4}{a^4 U''(a) \over r^2 U(r)}j_{\eps}(r) \cr A_\rho(r)& = -{1\over r U(r)}-{1\over 4}{a^4 U''(a) \over r^2 U(r)}j_\rho(r)~.        }}

In terms of these functions, the contributions to the stress tensor $T_2^{ij}$ become
\eqn\ogg{\eqalign{&[T_2^{ij}]^{st}= \cr &
-\int_{\infty}^a dr \Bigg\{\left({\rho^2 -B^2}\right) \left(- 2{j_{\eps} j_{\eps}' \over r}+2{j_{\eps} j_{\eps} \over r^2} +{1\over 2} j_{\eps}' j_{\eps}' \right)+ 2U A_{\eps}^2 +4B^2{A_{\eps} j_{\eps} \over r^2 }\Bigg\}[\p_i \eps \p_j \eps]^{st}
\cr &
+\int_{\infty}^a dr \Bigg\{(\rho^2 -B^2) \left(- 2{j_\rho j_\rho' \over r}+2{j_\rho j_\rho \over r^2} +{1\over 2} j_\rho' j_\rho' \right)+ 2U A_\rho^2 +4B^2{A_\rho j_\rho \over r^2 } +4{A_\rho \over r}\Bigg\}
[\p_i \rho \p_j \rho]^{st}
\cr &
- \int_{\infty}^a dr \Bigg\{(\rho^2 -B^2) \left(-2 {(j_{\eps} j_\rho' +j_{\eps}' j_\rho)\over r}+4 {j_{\eps} j_\rho \over r^2} +{j_{\eps}' j_\rho' }   \right) +4 U A_{\eps} A_\rho +4{A_{\eps} \over r}
\cr &
\quad\quad\quad\quad\quad\quad+4B^2 {(A_{\eps} j_\rho +A_\rho j_{\eps})\over r^2}\Bigg\}[\epsilon_{ik} \p_k \eps \p_j \rho]^{st}
\cr &
- {4 \rho B } \int_{\infty}^a dr~   \left(  {j_{\eps} j_{\eps}'\over r} -{j_{\eps} j_{\eps} \over r^2}-{1\over 4}{j_{\eps}' j_{\eps}'}+ {A_{\eps} j_{\eps} \over r^2}   \right) [\epsilon_{ik}\p_k \eps \p_j \eps]^{st}
\cr &
+4\rho B\int_{\infty}^a dr~  \left(  {j_\rho j_\rho'\over r} -{j_\rho j_\rho \over r^2}-{1\over 4}{j_\rho' j_\rho'}+ {A_\rho j_\rho \over r^2}   \right) [\epsilon_{ik}\p_k \rho  \p_j \rho   ]^{st}
\cr &
+ 4 \rho B \int_{\infty}^a dr~  \left({(j_{\eps} j_\rho' + j_{\eps}'j_\rho)\over r}- 2 {j_{\eps} j_\rho \over r^2} -{1\over 2}{j_{\eps}' j_\rho'}+{A_{\eps} j_\rho+ A_\rho j_{\eps}  \over r^2}   \right)[\p_i \eps \p_j \rho]^{st}~.   }}
Adding these to the terms in \bw, we can read off all of the second order nonlinear transport coefficients,
\eqn\ogh{\eqalign{
b_3^{\eps \eps} = &{1\over 4} {\p \over \p \eps}\left({a(\eps,\rho)^2 \over \rho^2+B^2}\right)
%
%\cr &
%
-\int_{\infty}^a dr \Bigg\{\left({\rho^2 -B^2}\right) \left(- 2{j_{\eps} j_{\eps}' \over r}+2{j_{\eps} j_{\eps} \over r^2} +{1\over 2} j_{\eps}' j_{\eps}' \right)+ 2U A_{\eps}^2 +4B^2{A_{\eps} j_{\eps} \over r^2 }\Bigg\} \cr
b_3^{\eps \rho} =& \left[ {1\over 4} {\p \over \p \rho}\left({a(\eps,\rho)^2 \over \rho^2+B^2}\right) -{1\over 2}  {\p \over \p \eps}\left({a(\eps,\rho)\rho \over \rho^2+B^2}\right)  \right] \cr
& + 4 \rho B \int_{\infty}^a dr~  \left({(j_{\eps} j_\rho' + j_{\eps}'j_\rho)\over r}- 2 {j_{\eps} j_\rho \over r^2} -{1\over 2}{j_{\eps}' j_\rho'}+{A_{\eps} j_\rho+ A_\rho j_{\eps}  \over r^2}   \right) \cr
b_3^{\rho \rho} =& -{1\over 2} {\p \over \p \rho}\left({a(\eps,\rho) \rho \over \rho^2+B^2}\right) \cr
& +\int_{\infty}^a dr \Bigg\{(\rho^2 -B^2) \left(- 2{j_\rho j_\rho' \over r}+2{j_\rho j_\rho \over r^2} +{1\over 2} j_\rho' j_\rho' \right)+ 2U A_\rho^2 +4B^2{A_\rho j_\rho \over r^2 } +4{A_\rho \over r}\Bigg\},
}}
and
\eqn\ogi{\eqalign{
b_4^{\eps \eps} =& {1 \over 4B}{\p \over \p \eps}\left({a(\eps,\rho)^2 \rho  \over \rho^2+B^2}\right)
%
%\cr &
%
- {4 \rho B } \int_{\infty}^a dr~   \left(  {j_{\eps} j_{\eps}'\over r} -{j_{\eps} j_{\eps} \over r^2}-{1\over 4}{j_{\eps}' j_{\eps}'}+ {A_{\eps} j_{\eps} \over r^2}   \right)
\cr
b_4^{\eps \rho} =& {1 \over 4B}{\p \over \p \rho}\left({a(\eps,\rho)^2 \rho  \over \rho^2+B^2}\right) +{B \over 2}{\p \over \p \eps}\left({a(\eps,\rho) \over \rho^2+B^2}\right)
\cr &
- \int_{\infty}^a dr \Bigg\{(\rho^2 -B^2) \left(-2 {(j_{\eps} j_\rho' +j_{\eps}' j_\rho)\over r}+4 {j_{\eps} j_\rho \over r^2} +{j_{\eps}' j_\rho' }   \right) +4 U A_{\eps} A_\rho +4{A_{\eps} \over r}
\cr &
\quad\quad\quad\quad\quad\quad+4B^2 {(A_{\eps} j_\rho +A_\rho j_{\eps})\over r^2}\Bigg\} \cr
b_4^{\rho \rho} =& {B \over 2}{\p \over \p \rho}\left({a(\eps,\rho) \over \rho^2+B^2}\right) +4\rho B\int_{\infty}^a dr~  \left(  {j_\rho j_\rho'\over r} -{j_\rho j_\rho \over r^2}-{1\over 4}{j_\rho' j_\rho'}+ {A_\rho j_\rho \over r^2}   \right).
}}

The final step in the construction of our second order solution  is to check that the remaining Einstein-Maxwell equations, $W^{(2)}_{Mv}=Y^{(2)r}=0$,
imply the equations of fluid dynamics  \A\ at order $\ve^2$.   This is indeed the case.

\newsec{Conclusion}

We have shown how the dynamics of a dyonic black brane can give rise to nonlinear magnetohydrodynamic equations of motion.  We explicitly solved the Einstein-Maxwell equations to second order in the fluid dynamical derivative expansion, and from this we found expressions for the stress tensor and current to second and third order in derivatives respectively.  These expressions yielded gravity predictions for the magnetohydrodynamic transport coefficients of charge and energy densities in a specific conformal fluid, coefficients which can in principle be compared with experimental results dealing with  2+1 dimensional fluids near a quantum critical point. Our linear transport coefficients  extend to higher order results already appearing in the literature, and a new set of nonlinear transport coefficients is now available as well. We found
explicit closed form expressions for the second order linear transport coefficients, and integral
expressions for the nonlinear ones.

Our work also helped to demonstrate the broad applicability of the mechanism  \NLFDFG\ by which the
equations of fluid dynamics emerge from gravity at long wavelengths.   Compared to previous examples, one novelty of our setup was that it did not involve
considering boosted black branes, due to the breaking of Lorentz invariance by the background magnetic
field.  Instead, we only needed to consider fluctuations of the energy and charge density.

Various generalizations and extensions of this work are possible, such as considering higher dimensional
spacetimes, curved boundaries, etc.  As with other examples, it may be interesting to consider in more
detail particular solutions, and to see what can be learned about gravity from properties of the fluid dynamics, and vice versa.

\bigskip
\noindent {\bf Acknowledgments:} \medskip \noindent  We thank Sean Hartnoll, Chris Herzog,  Esko Keski-Vakkuri, and Markus Muller for helpful discussions and correspondence.     Work of PK is supported in part by NSF grant PHY-0456200.

\appendix{A}{Integrals}

Here we give the results for the integrals appearing in \bg.
First write $U$ in terms of its roots as $U = {1\over r^2}(r-r_1)(r-r_2)(r-r_3)(r-r_4)$.  We note that one of the roots is equal to $a$, while the other three can be obtained explicitly as the roots of
a cubic equation.

We will use the  identity
\eqn\za{ {1 \over U^2} = \sum_i{1\over [U'(r_i)]^2}{1\over (r-r_i)^2} - \sum_i{U''(r_i)\over [U'(r_i)]^3}{1\over (r-r_i)} }
along with
\eqn\zb{\eqalign{ &\sum_i \left( {1\over [U'(r_i)]^2} {1\over r_i^2} + {U''(r_i)\over [U'(r_i)]^3}{1\over r_i}\right)=0 \cr & \sum_i \left(2 {1\over [U'(r_i)]^2} {1\over r_i^3} + {U''(r_i)\over [U'(r_i)]^3}{1\over r_i^2}\right)=0~. }}
The identities in \zb\ following from applying derivatives to \za, evaluated at $r=0$.  We then find

\eqn\zc{\eqalign{ \int\! {dr \over U^2} & = -\sum_i{1\over [U'(r_i)]^2}{1\over (r-r_i)} - \sum_i{U''(r_i)\over [U'(r_i)]^3}\ln(r-r_i) \cr  \int\! {dr \over r U^2} & =-\sum_i{1\over [U'(r_i)]^2}\left({1\over r_i (r-r_i)}+{\ln(r-r_i) \over r_i^2 } \right)- \sum_i{U''(r_i)\over [U'(r_i)]^3}{\ln(r-r_i)\over r_i} \cr  \int\! {dr \over r^2 U^2} & = -\sum_i{1\over [U'(r_i)]^2}\left( {1\over r_i^2 (r-r_i)} +2 {\ln(r-r_i)\over r_i^3}\right) - \sum_i{U''(r_i)\over [U'(r_i)]^3}{\ln(r-r_i)\over r_i^2}~. }}

\listrefs
\end